\documentclass{myPoS}
\usepackage{amsmath} 
\usepackage{amssymb}
\usepackage{mathrsfs}
\usepackage{fontenc} 
\usepackage{graphicx}
\usepackage{graphics}
\usepackage{mathptmx}
\usepackage{times} 

\newcommand{\be}{\begin{equation}}
\newcommand{\ee}{\end{equation}}
\newcommand{\ba}{\begin{eqnarray}}
\newcommand{\ea}{\end{eqnarray}}
\def\beann{\begin{eqnarray*}} \def\eeann{\end{eqnarray*}}
\newcommand{\bal}{\begin{align}}
\newcommand{\eal}{\end{align}}

\def\lsim{\raise0.3ex\hbox{$<$\kern-0.75em\raise-1.1ex\hbox{$\sim$}}}
\def\gsim{\raise0.3ex\hbox{$>$\kern-0.75em\raise-1.1ex\hbox{$\sim$}}}

\title{
\vspace*{-1.5cm}
\begin{flushright}\texttt{\footnotesize
CERN-PH-TH/2012-003 \\ YITP-12-2 \\ \vspace*{-0.36cm} IFUP-TH/2012-01 }
\end{flushright}
\vfill
Constraints on the two-flavor QCD phase diagram \\
from imaginary chemical potential
}

\ShortTitle{$N_f=2$ phase diagram} 

\author{Claudio Bonati \\
Dipartimento di Fisica, Universit\`a di Pisa and INFN, \\
Sezione di Pisa, Largo Pontecorvo 3, 56127 Pisa, Italy \\
E-mail: \email{bonati@df.unipi.it}}

\author{\speaker{Philippe de Forcrand}\thanks{Temporary address:
Yukawa Institute for Theoretical Physics, Kyoto University, 
Kyoto 606-8502, Japan} \\ 
Institute for Theoretical Physics, ETH Z\"urich, CH-8093 Z\"urich, Switzerland \\
and \\
CERN, Physics Department, TH Unit, CH-1211 Geneva 23, Switzerland \\ 
E-mail: \email{forcrand@phys.ethz.ch}} 

\author{Massimo D'Elia \\
Dipartimento di Fisica, Universit\`a di Pisa and INFN, \\
Largo Pontecorvo 3, 56127 Pisa, Italy \\
E-mail: \email{delia@df.unipi.it}}

\author{Owe Philipsen \\
        Institut f\"ur Theoretische Physik, Goethe-Universit\"at Frankfurt, 60438
Frankfurt am Main, Germany \\
        E-mail: \email{philipsen@th.physik.uni-frankfurt.de}}
        
\author{Francesco Sanfilippo \\
Laboratoire de Physique Th\'eorique (B\^at.~210) \footnote{Laboratoire de
Physique Th\'eorique est une unit\'e mixte de recherche du CNRS, UMR 8627.} 
Universit\'e  Paris Sud, Centre d'Orsay, F-91405
Orsay-Cedex, France, and \\
INFN Sezione di Roma, P.le Aldo Moro 5, 00185 Roma, Italy \\
E-mail: \email{francesco.sanfilippo@th.u-psud.fr}}

\abstract{
We review our knowledge of the phase diagram of QCD as a function of 
temperature, chemical potential and quark masses. 
The presence of tricritical lines at imaginary chemical potential 
$\mu=i\frac{\pi}{3}T$, with known scaling behaviour in their vicinity, 
puts constraints on this phase diagram, especially in the case of two light 
flavors. We show first results in our project to determine the 
finite-temperature behaviour in the $N_f=2$ chiral limit.
} 

\FullConference{ The XXIX International Symposium on Lattice Field Theory - Lattice 2011\\
July 10-16, 2011\\
Squaw Valley, Lake Tahoe, California}

\begin{document} 

\section{Introduction}

The phase diagram of QCD as a function of temperature $T$ and quark chemical potential $\mu$
is governed by the interplay of the chiral symmetry and the center symmetry~\cite{reviewOP}. These symmetries are exact for zero and infinite quark masses, respectively. Therefore, varying the quark masses
away from their physical values towards these limits provides useful insight into the behaviour
of QCD at the physical mass point.

Similarly, making the chemical potential complex provides enhanced information on the behaviour
of QCD at real chemical potential. Since the sign problem which plagues simulations at non-zero
$\mu$~\cite{reviewPdF} is absent when $\mu$ is pure imaginary, the regime of imaginary 
$\mu$ is actually the only direction in the complex $\mu$ plane where complete, reliable information
on the behaviour of QCD can be obtained. It turns out that a rich phase diagram as a function
of $(T,\mu=i\mu_I)$ and of the quark masses $(m_u=m_d,m_s)$ emerges. The critical and tricritical
features of this phase diagram, with their associated scaling laws, have consequences for the 
behaviour of QCD at real $\mu$.

Here, we summarize what is known about this phase diagram and sketch (Fig.~2) a plausible 
scenario, consistent with current numerical simulations augmented with reasonable assumptions of
continuity of the critical surfaces. We explore in particular the implications for the behaviour
of QCD in the two-flavor chiral limit $(m_u=m_d=0,m_s=\infty)$. In that limit, it is widely believed that QCD undergoes a finite-temperature, second-order $O(4)$ chiral transition at $\mu=0$, which turns
first-order at a tricritical point for some real $\mu$ \cite{theory}. However, other possibilities exist.
At $\mu=0$ in particular, the finite-temperature transition might be first-order. The present numerical evidence is inconclusive: using Wilson fermions, $O(4)$ scaling is preferred \cite{Tsukuba},
while with staggered fermions $O(4)$ scaling has been elusive, and first-order behaviour has also
been claimed \cite{DiG}. Note that behaviour consistent with $O(4)$ has been seen with improved 
staggered fermions, in an $N_f=2+1$ setup where the strange quark mass is fixed at its physical value \cite{Unger}.
Approaching the chiral limit from the imaginary $\mu$ direction offers a novel, independent method
to help settle the issue.

\section{Three-dimensional Columbia plot}

The thermal behaviour of QCD at $\mu=0$, as a function of the quark masses $m_u=m_d\equiv m_{u,d}$ and
$m_s$ is summarized in the well-known Columbia plot Fig.~1 ({\em left}). The $N_f=3$ chiral symmetry and the $Z_3$ center symmetry are achieved in the lower left and upper right corners, respectively. This
gives rise to first-order transitions. For intermediate quark masses, numerical simulations indicate 
a smooth crossover as a function of temperature. Hence, the first-order regions must be bounded by
second-order critical lines: the chiral critical line in the lower left corner, and the deconfinement
critical line in the upper right corner. In the absence of further symmetry, the universality class
is expected (and has been numerically verified) to be that of the $3d$ Ising model.
The chiral critical line joins with the $m_{u,d}=0$ axis at a tricritical point, for a strange quark
mass $m_s^{\rm tric}$ which
is larger than the physical strange quark mass on coarse lattices \cite{OP2} 
or smaller when using improved actions \cite{Unger}. 
The $N_f=2$ chiral limit is obtained in the upper left corner.

When the chemical potential is turned on, the two critical lines sweep critical surfaces as a function
of $\mu$. For both lines, it has been observed that the first-order region shrinks, as represented 
Fig.~1 ({\em right})~\cite{OP2,OP3,Potts}. Here, we want to show real and imaginary $\mu$ in a single
figure. Therefore, we adopt $(\mu/T)^2$ for the $z$ coordinate: real and imaginary $\mu$ appear above
and below the $\mu=0$ plane, respectively. Of particular interest is the Roberge-Weiss plane
$(\mu/T)^2 = -(\pi/3)^2$. We now argue that the 3-dimensional phase diagram of $N_f=2+1$ QCD is likely
to be described by Fig.~2.

\begin{figure}[t]
\begin{center}
\centerline{
\includegraphics[width=0.35\textwidth]{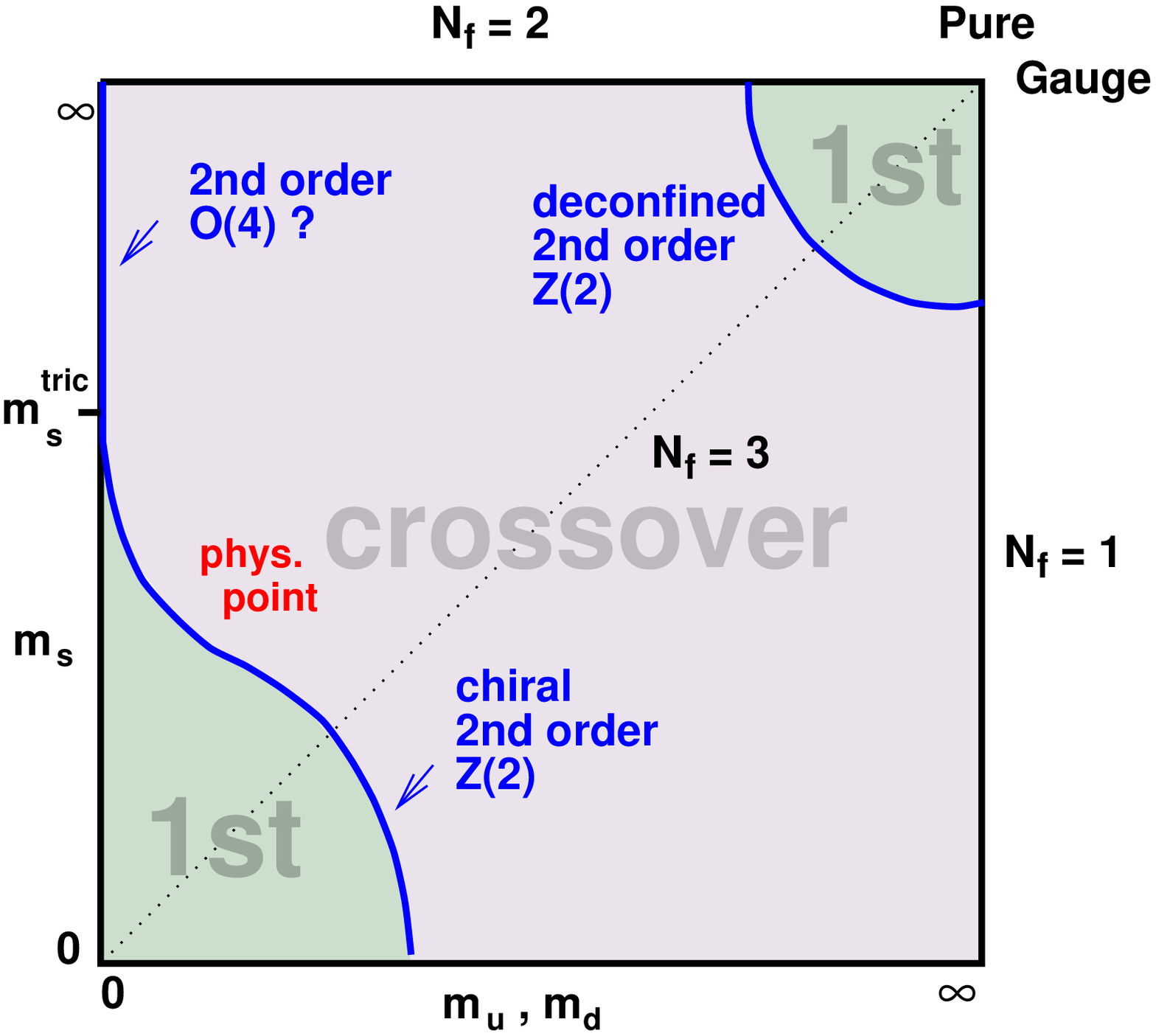}
\hspace*{0.5cm}
\includegraphics[width=0.55\textwidth]{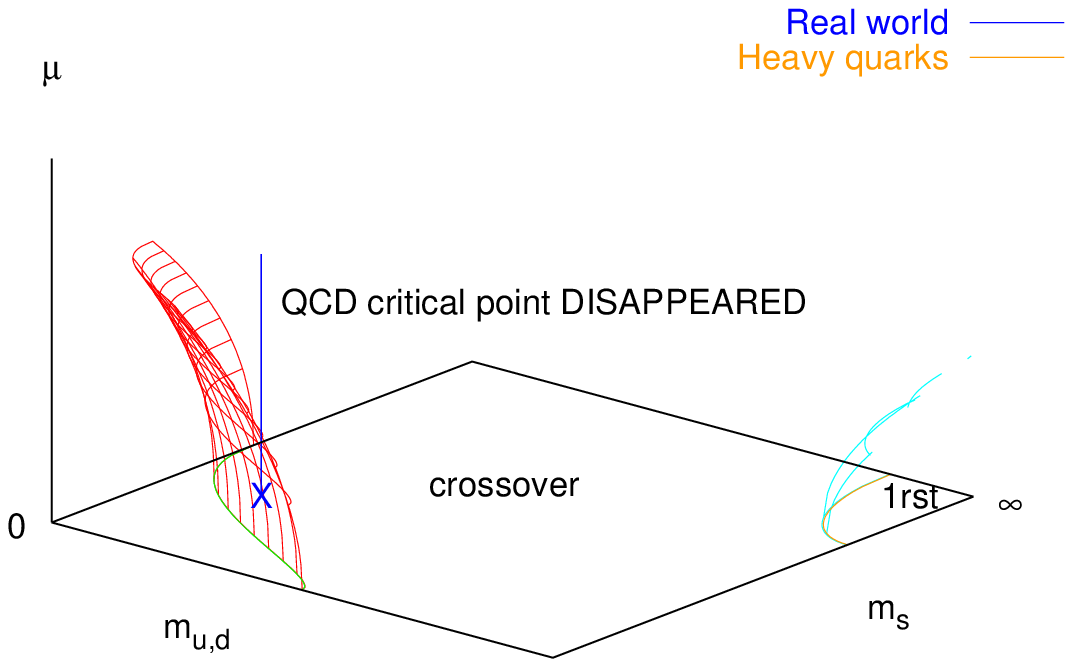}
}
\caption{({\em Left}) ``Columbia plot'': schematic phase transition behaviour of $N_f=2+1$ QCD
for different choices of quark masses $(m_{u,d},m_s)$ at $\mu=0$. Two critical lines separate
the regions of first-order transitions (light or heavy quarks) from the crossover region in the middle,
which includes the physical point.
({\em Right}) Critical surfaces swept by the critical lines as $\mu$ is turned on. 
For light quarks~\cite{OP2,OP3,OP1} as well as for heavy quarks~\cite{Potts}, numerical simulations indicate that the first-order region shrinks as the chemical potential is turned on.
}
\end{center}
\vspace*{-0.5cm}
\end{figure}

\begin{figure}[h]
\begin{center}
\centerline{
\includegraphics[width=1.00\textwidth]{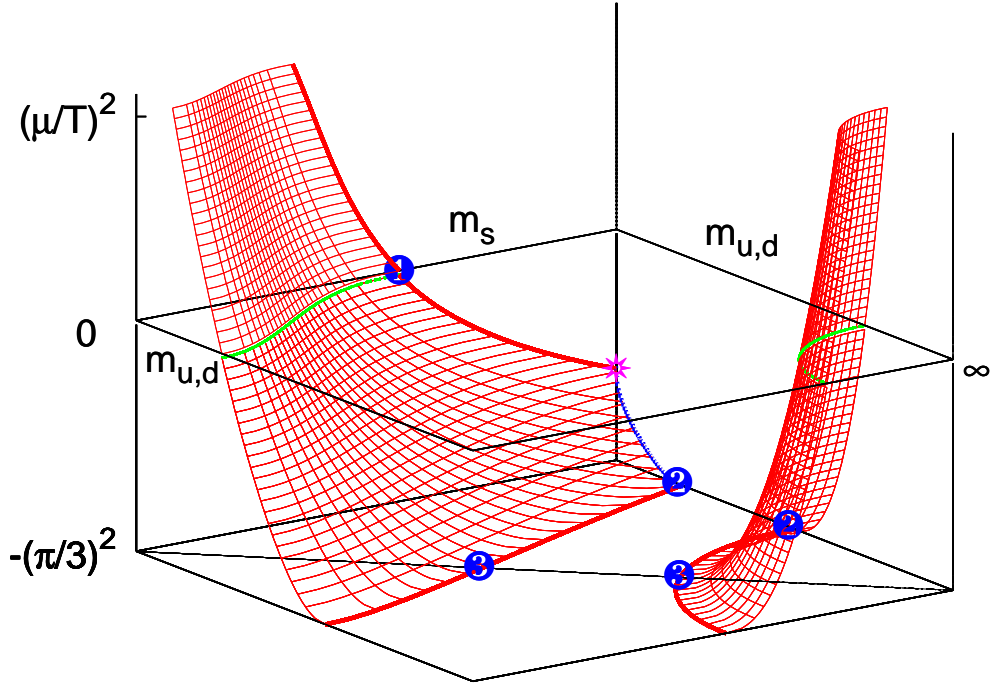}
}
\caption{$3d$ phase diagram. The vertical axis is $(\mu/T)^2$, so that real
and imaginary chemical potentials are above and below the $\mu=0$ plane,
respectively. The ``bottom plane'' corresponds to the Roberge-Weiss transition value
$\mu/T=i\pi/3$. The thicker red lines are tricritical. Tricritical points marked ``2'' and ``3''
have been identified for the $N_f=2$~\cite{DEliaRW} and $N_f=3$~\cite{OPRW} theories, respectively.
The object of the present study is the blue line in the ``backplane'' $m_s=\infty$ ($N_f=2$)
joining two tricritical points.
}
\end{center}
\vspace*{-0.5cm}
\end{figure}

\section{Phase diagram in the Roberge-Weiss plane}

\begin{figure}[t]
\begin{center}
\centerline{
\includegraphics[width=0.318\textwidth]{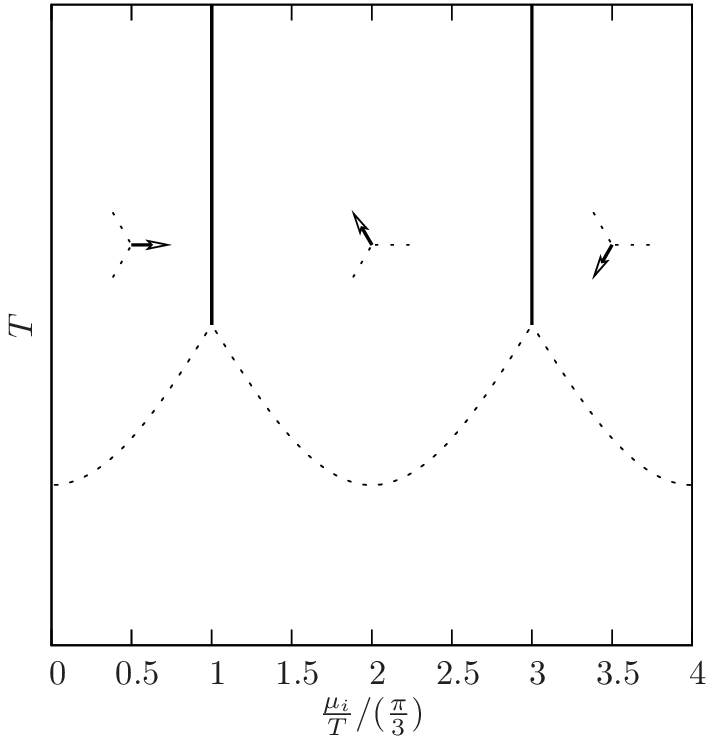}
\hspace*{0.5cm}
\includegraphics[width=0.32\textwidth]{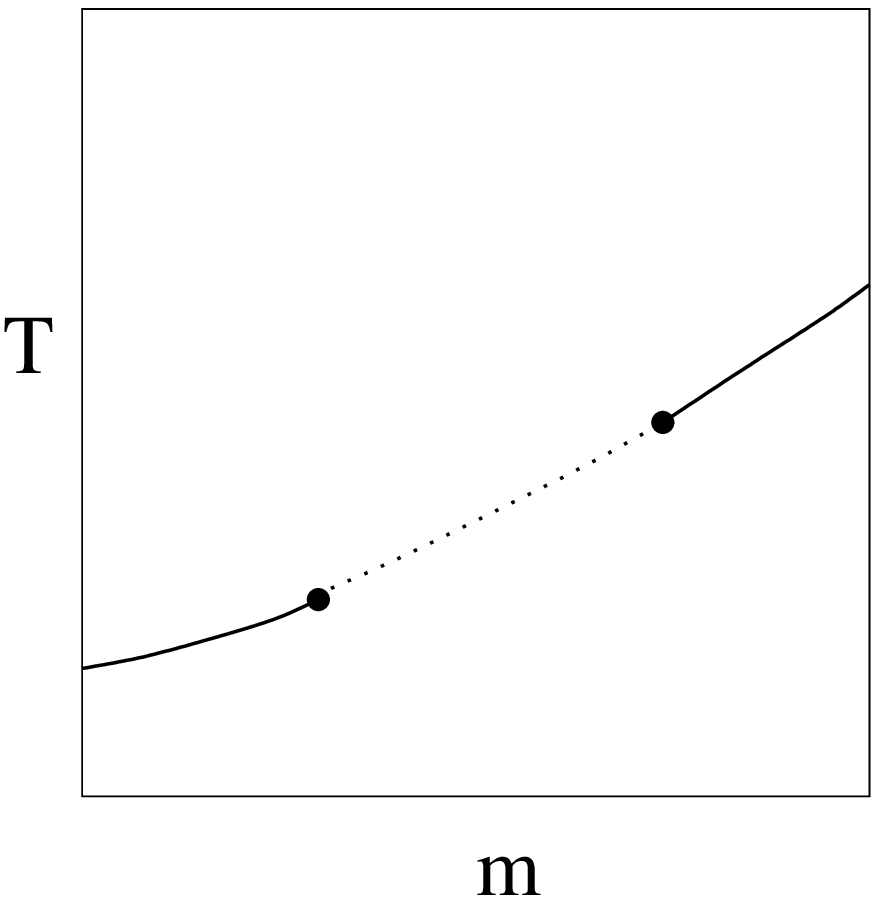}
\put(-64,83){\color{red}\small Tricritical}
\put(-97,35){\color{red}\small Tricritical}
\hspace*{0.5cm}
\includegraphics[width=0.32\textwidth]{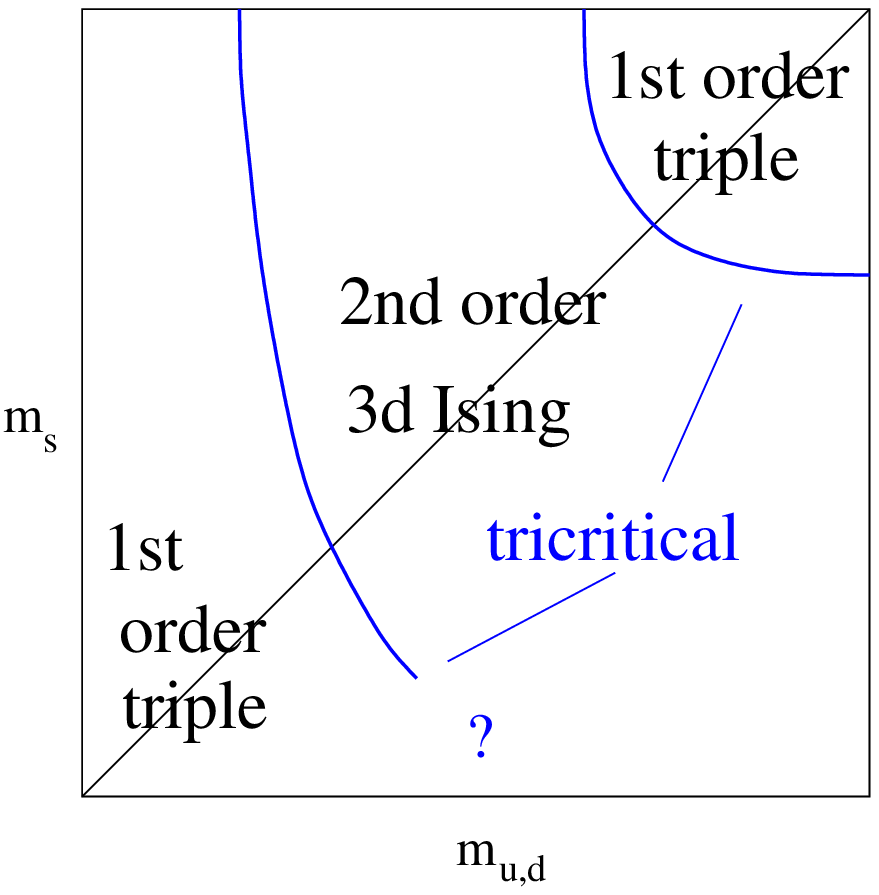}
}
\caption{({\em Left}) Generic phase diagram as a function of imaginary chemical potential and temperature. Solid lines are first-order Roberge-Weiss transitions. The behaviour along dotted lines depends on the number of flavors and the quark masses. ({\em Middle}) For $N_f=2$ and $N_f=3$, the endpoint of the Roberge-Weiss line is a triple point (where 3 first-order lines meet) for light or heavy quark masses,
and an Ising critical point for intermediate quark masses. Thus, two tricritical masses exist. 
({\em Right}) The simplest assumption is that the $N_f=2$ and $N_f=3$ tricritical points are joined by
tricritical lines~\cite{OPRW}. This assumption can be checked with $N_f=2+1$ imaginary-$\mu$ simulations.
}
\end{center}
\vspace*{-0.5cm}
\end{figure}

The two symmetries of the partition function
\be
Z(\mu)=Z(-\mu), ~~~ Z\left(\frac{\mu}{T}\right) = Z\left(\frac{\mu}{T} + i\frac{2\pi n}{3}\right)
\ee
imply reflection symmetry in the imaginary $\mu$ direction about the ``Roberge-Weiss'' values $\mu = i\pi T/3 (2n+1)$ which separate different sectors of
the center symmetry~\cite{RW}.
Transitions between neighbouring sectors
are of first order for high $T$ and analytic crossovers
for low $T$ \cite{RW,fp1,el1},
as indicated Fig.~3 ({\em left}). 
The corresponding first-order transition lines may end
with a second-order critical point, or with a triple point, branching off into two first-order lines.
Which of these two possibilities occurs depends on the number of flavors and the quark masses. 

Recent numerical studies have shown that a triple point is found for heavy and light quark masses,
while for intermediate masses one finds a second-order endpoint. As a function of the quark mass, the phase diagram at $\mu/T = i\pi/3$ is as sketched Fig.~3 ({\em middle}).
This happens for both $N_f=2$ \cite{DEliaRW} and $N_f=3$ \cite{OPRW}. 

If one assumes that the $N_f=2$ and $N_f=3$ tricritical points are connected to each other in the
$(m_{u,d},m_s)$ quark mass plane, the resulting phase diagram is depicted Fig.~3 ({\em right}),
with two tricritical lines separating regions of first-order and of second-order transitions.
This phase diagram is the equivalent of the Columbia plot, now at imaginary chemical potential
$\mu/T = i\pi/3$. Note that the assumption of continuity of the tricritical lines can be checked
directly 
by numerical simulations, since there is no sign problem for imaginary $\mu$.

Now, as $(\mu/T)^2$ is varied between zero and the Roberge-Weiss value $-(\pi/3)^2$, the Columbia plot
must change from Fig.~1 ({\em left}) to Fig.~3 ({\em right}). Assuming continuity of the critical surfaces at imaginary $\mu$, which again can be checked by numerical simulations, the resulting 3-dimensional
phase diagram is that of Fig.~2. The two red surfaces (``chiral'' and ``deconfinement'') are critical. They are bounded by lines, among which the following are tricritical: ({\em i}) the two lines in the $(\mu/T)^2=-(\pi/3)^2$ Roberge-Weiss plane; ({\em ii}) the line in the $m_{u,d}=0$ chiral plane. Note that the $N_f=2$ (i.e. $m_s=\infty$) ``backplane'' contains two tricritical points on the chiral critical surface: one in the Roberge-Weiss plane, the other on the $m_{u,d}=0$ vertical axis (see Fig.~2). The location of the latter is related to the value of the tricritical strange quark mass.

\section{Tricritical scaling}

In the vicinity of a tricritical point, scaling laws apply. The phase diagram is similar to that of 
a metamagnet, with two external fields: $H$, which respects the symmetry, and $H^\dagger$ which breaks it (like a staggered and an ordinary magnetic field), depicted Fig.~4 ({\em left}).
The three surfaces $S_0, S_+, S_-$ indicate first-order transitions. They meet at a line of triple points $L_\tau$, depicted by a solid line. They are bounded by second-order transition lines, depicted by dotted lines. All four lines meet at the tricritical point $(T_t,H_t)$.

In our case, the scaling exponents governing the behaviour near the tricritical point are mean-field,
because QCD becomes 3-dimensional as the correlation length diverges while the temperature is fixed
to that of the tricritical point, and $d=3$ is the upper critical dimension for tricriticality.
Of course, this implies the presence of potentially large logarithmic corrections to scaling.

Here, we are interested in the second-order lines $S_\pm$, corresponding to a departure from the symmetry plane $H^\dagger=0$. Along these lines, the scaling law is $H^\dagger \propto |t|^{5/2}$, where the
reduced temperature $t$ is measured along the tangent to $L_\lambda$. For $N_f=2$ QCD, tricritical scaling
should be satisfied near the tricritical points: \\
$(i)$ $(\mu/T)^2 = -(\pi/3)^2$: then $H^\dagger \sim \left[(\mu/T)^2 + (\pi/3)^2\right], ~~ t \sim (m_{u,d}-m_{\rm tric})$, so that
\be
\left[(\mu/T)^2 + (\pi/3)^2\right] \propto (m_{u,d}-m_{\rm tric})^{5/2}
\label{eq1}
\ee
$(ii)$ $m_{u,d}=0$: then $H^\dagger \sim m_{u,d}, ~~ t \sim \left[(\mu/T)^2 - (\mu/T)^2|_{\rm tric}\right]$, so that
\be
m_{u,d} \propto \left[(\mu/T)^2 - (\mu/T)^2|_{\rm tric}\right]^{5/2}
\label{eq2}
\ee
It is not clear how broad the scaling window is around each of these two tricritical points. The two scaling windows might overlap, leading to a very constrained system of equations, with 3 unknowns (the two constants of proportionality in eqs.(\ref{eq1},\ref{eq2}) and $(\mu/T)^2|_{\rm tric}$ --- $m_{\rm tric}$ having been determined already in \cite{DEliaRW}) and one constraint (continuity of the derivative at the intersection of the two scaling curves), leading to a phase diagram as in Fig.~5 ({\em left}), which could be determined from only two
points measured by Monte Carlo. Reason for such optimism can be found in Fig.~4 ({\em right}), where the
scaling window around the third $N_f=2$ tricritical point, corresponding to heavy $(u,d)$ quarks, is
shown to extend far into the region of real chemical potential~\cite{OPRW,opqcd}.

\begin{figure}[t]
\begin{center}
\centerline{
\includegraphics[width=0.32\textwidth]{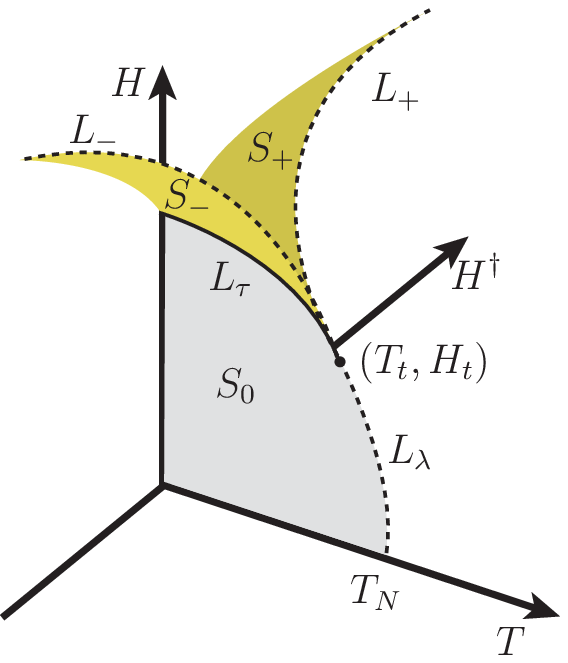}
\hspace*{1.0cm}
\includegraphics[width=0.50\textwidth]{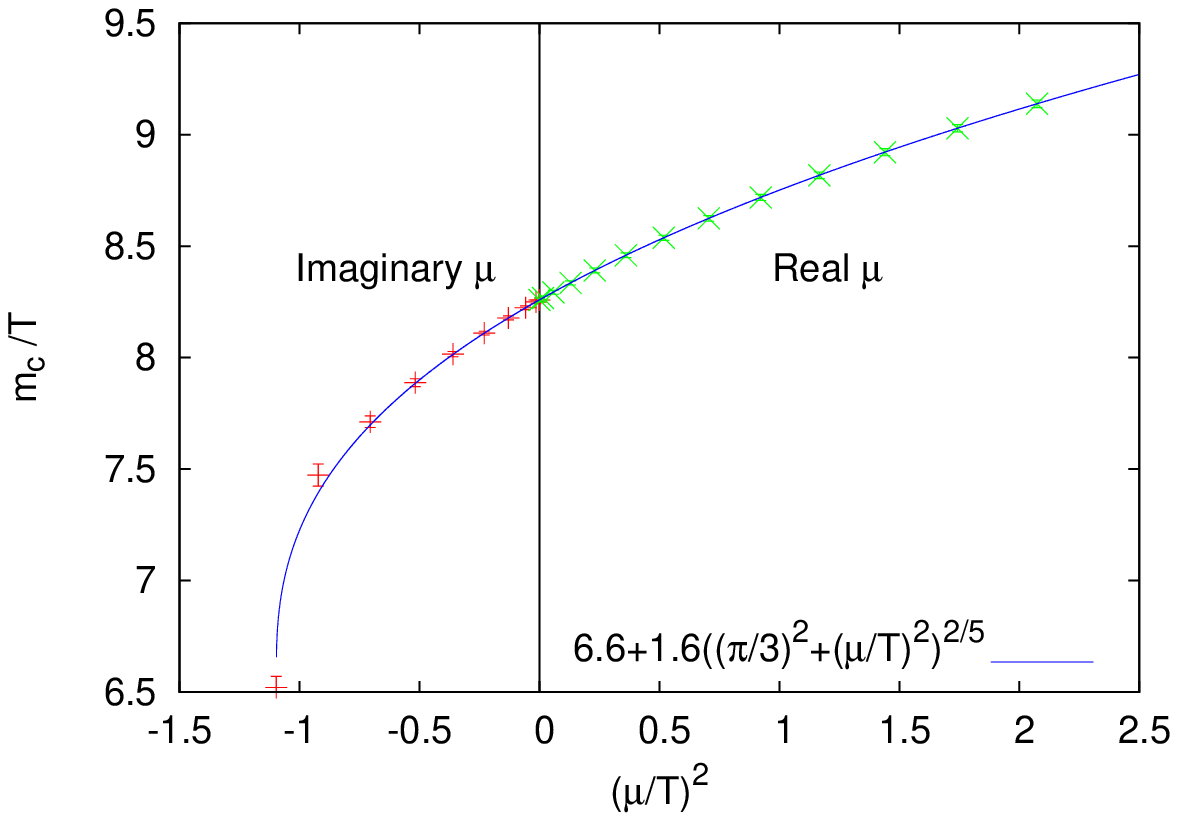}
\put(-161.0,26.0){\color{red}$\bullet$} 
}
\caption{(Left) Schematic phase diagram of a metamagnet. The external field $H^\dagger$ breaks the
symmetry, while $H$ does not.
(Right) For heavy quarks, tricritical scaling in the vicinity of the
Roberge-Weiss imaginary-$\mu$ value extends far into the region of real $\mu$~\cite{OPRW}.
}
\end{center}
\vspace*{-0.5cm}
\end{figure}

\section{Preliminary $N_f=2$ results}

Following the above discussion, we performed simulations of $N_f=2$ QCD, with staggered quarks of masses
$a m_q = 0.01$ and $0.005$ on $N_t=4$ lattices, scanning in $(\mu/T)^2$ to determine the value of imaginary $\mu$ corresponding to a second-order transition. Our observable is the Binder cumulant of the quark condensate. Consistent results are obtained from the finite-size scaling of the plaquette distribution.
The two critical points are shown Fig.~5 ({\em right}). Disappointingly, it seems impossible to smoothly
match two tricritical scaling curves passing through these points. Additional masses are needed to determine the critical curve. Nevertheless, assuming convexity of the critical curve already
constrains the $m_{u,d}=0$ tricritical point to lie at $(\mu/T)^2 \gtrsim -0.3$. The figure illustrates
the case where this point lies at $\mu=0$. It might also lie at $(\mu/T)^2 > 0$, so that the $\mu=0$
chiral transition would be first-order. Additional small-mass measurements are underway and will settle this issue. Note that we are simulating two-flavour QCD by taking the square root of the staggered determinant, and approaching the chiral limit at fixed, rather coarse lattice spacing. This is the wrong
order of limits, and is the most likely approach to expose a failure of rooting. 

\begin{figure}[t]
\begin{center}
\centerline{
\includegraphics[width=0.55\textwidth]{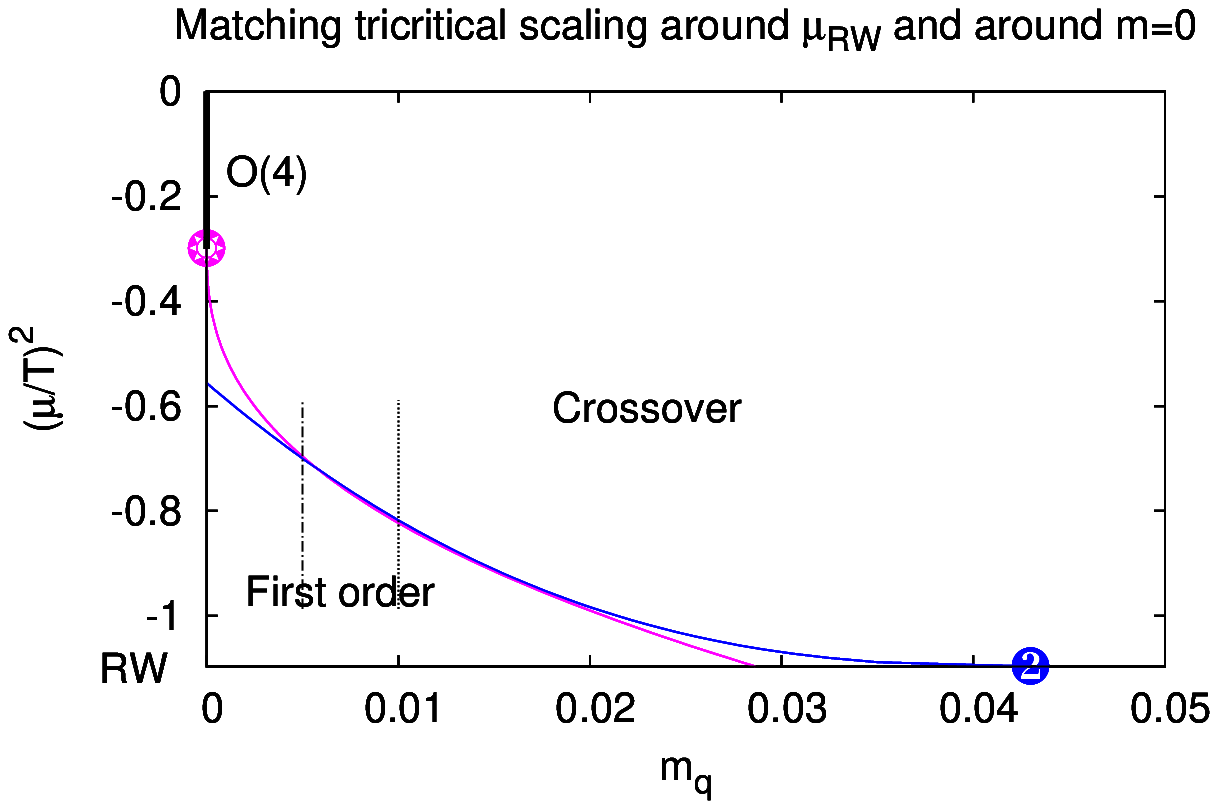}
\hspace*{0.5cm}
\includegraphics[width=0.55\textwidth]{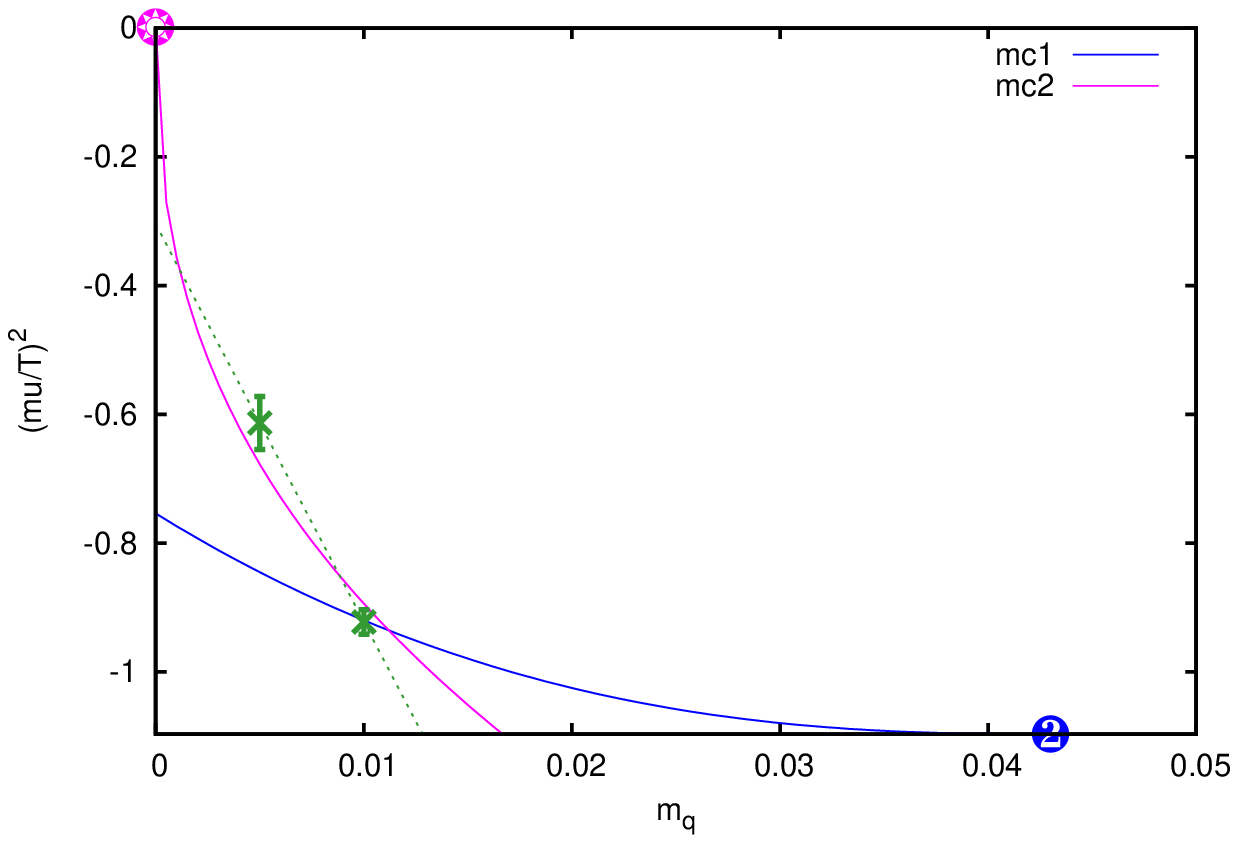}
}
\caption{({\em Left}) Strategy used: to determine the critical line, we fix
the quark mass and measure the corresponding critical imaginary chemical 
potential. The two tricritical scaling lines may match smoothly if the scaling windows overlap.
({\em Right}) Preliminary results: the scaling windows do not seem to overlap; convexity places
the $m_{u,d}=0$ tricritical point at $(\mu/T)^2 \gtrsim -0.3$. For illustration, it is placed at 
$\mu=0$ in the figure.
}
\end{center}
\vspace*{-0.5cm}
\end{figure}

Finally, our phase diagram Fig.~2 makes it clear that, if the transition in
the massless $N_f = 2$ theory is $O(4)$, turning on a real chemical potential
(i.e. going up the vertical axis in the back)
will not make it become first-order. Obtaining such behaviour requires that
the chiral critical surface bend away from the $N_f=3$ chiral point,
or that another, non-chiral critical surface appear at large \nolinebreak $\mu$.

\section*{Acknowledgements:}
This work is partially supported by the German BMBF, project
{\em Hot Nuclear Matter from Heavy Ion Collisions
and its Understanding from QCD}, No.~06MS254.
We thank the Minnesota Supercomputer Institute and HLRS Stuttgart
for providing computer resources.
Large scale simulations
have been performed on two GPU clusters in Pisa and
Genoa provided by INFN.


\begin{thebibliography}{99}

\bibitem{reviewOP}
For a review, see e.g. 
  E.~Laermann and O.~Philipsen,
  Ann.\ Rev.\ Nucl.\ Part.\ Sci.\  {\bf 53} (2003) 163
  [hep-ph/0303042].

\bibitem{reviewPdF}
  P.~de Forcrand,
  PoS {\bf LAT2009} (2009) 010
  [arXiv:1005.0539 [hep-lat]].

\bibitem{theory}
A.~M.~Halasz, A.~D.~Jackson, R.~E.~Shrock, M.~A.~Stephanov and J.~J.~M.~Verbaarschot,
  Phys.\ Rev.\ D {\bf 58} (1998) 096007
  [hep-ph/9804290].

\bibitem{Tsukuba}
  Y.~Iwasaki, K.~Kanaya, S.~Kaya and T.~Yoshie,
  Phys.\ Rev.\ Lett.\  {\bf 78} (1997) 179
  [hep-lat/9609022].

\bibitem{DiG}
M.~D'Elia, A.~Di Giacomo and C.~Pica,
  Phys.\ Rev.\ D {\bf 72}, 114510 (2005)
  [hep-lat/0503030];
G.~Cossu, M.~D'Elia, A.~Di Giacomo and C.~Pica,
  arXiv:0706.4470 [hep-lat];
  C.~Bonati, G.~Cossu, M.~D'Elia, A.~Di Giacomo and C.~Pica,
  PoS LATTICE {\bf 2008} (2008) 204
  [arXiv:0901.3231 [hep-lat]].

\bibitem{Unger}
  S.~Ejiri, F.~Karsch, E.~Laermann, C.~Miao, S.~Mukherjee, P.~Petreczky, C.~Schmidt and W.~Soeldner {\it et al.},
  Phys.\ Rev.\ D {\bf 80} (2009) 094505
  [arXiv:0909.5122 [hep-lat]].

\bibitem{OP2}
  P.~de Forcrand and O.~Philipsen,
  JHEP {\bf 0701} (2007) 077
  [arXiv:hep-lat/0607017].

\bibitem{OP3}
  P.~de Forcrand and O.~Philipsen,
  JHEP {\bf 0811} (2008) 012
  [arXiv:0808.1096 [hep-lat]].

\bibitem{Potts}
  S.~Kim et al.,
  PoS {\bf LAT2005} (2006) 166
  [arXiv:hep-lat/0510069].

\bibitem{OP1}
  P.~de Forcrand and O.~Philipsen,
  Nucl.\ Phys.\  B {\bf 673} (2003) 170
  [arXiv:hep-lat/0307020].
  
\bibitem{RW}
  A.~Roberge and N.~Weiss,
  Nucl.\ Phys.\  B {\bf 275} (1986) 734.

\bibitem{fp1}
P.~de Forcrand and O.~Philipsen,
  Nucl.\ Phys.\  B {\bf 642}, 290 (2002)
  [arXiv:hep-lat/0205016].

\bibitem{el1}
M.~D'Elia and M.~P.~Lombardo,
  Phys.\ Rev.\  D {\bf 67}, 014505 (2003)
  [arXiv:hep-lat/0209146].

\bibitem{DEliaRW}
  M.~D'Elia and F.~Sanfilippo,
  Phys.\ Rev.\  D {\bf 80} (2009) 111501
  [arXiv:0909.0254 [hep-lat]];
C.~Bonati, G.~Cossu, M.~D'Elia and F.~Sanfilippo,
  Phys.\ Rev.\ D {\bf 83}, 054505 (2011)
  [arXiv:1011.4515 [hep-lat]].

\bibitem{OPRW}
  P.~de Forcrand and O.~Philipsen,
  Phys.\ Rev.\ Lett.\  {\bf 105} (2010) 152001
  [arXiv:1004.3144 [hep-lat]].

\bibitem{opqcd}
M.~Fromm, J.~Langelage, S.~Lottini and O.~Philipsen,
  arXiv:1111.4953 [hep-lat].


\end{thebibliography}
\end{document}